\documentclass[lettersize,journal]{IEEEtran}
\usepackage{amsmath,amsfonts}
\usepackage{algorithmic}
\usepackage{array}
\usepackage[caption=false,font=normalsize,labelfont=rm,textfont=rm]{subfig}
\usepackage{textcomp}
\usepackage{stfloats}
\usepackage{url}
\usepackage{verbatim}
\usepackage{graphicx}
\def\BibTeX{{\rm B\kern-.05em{\sc i\kern-.025em b}\kern-.08em
		T\kern-.1667em\lower.7ex\hbox{E}\kern-.125emX}}
\usepackage{balance}
\usepackage{algorithm}
\usepackage{color}
\usepackage{multirow}
\usepackage{hyperref} 
\hypersetup{
	colorlinks=true,
	linkcolor=blue,
	citecolor=blue,
	urlcolor=blue
}
\usepackage{mathrsfs}
\usepackage{cite}
\usepackage{booktabs}
\usepackage{adjustbox}
\begin{document}

\title{FSAR-Cap: A Fine-Grained Two-Stage Annotated Dataset for SAR Image Captioning}

\author{Jinqi Zhang, Lamei Zhang,~\IEEEmembership{Senior Member,~IEEE} and Bin Zou,~\IEEEmembership{Senior Member,~IEEE}

\thanks{This work was supported in part by the National Natural Science Foundation of China (62271172). \textit{(Corresponding author: Lamei Zhang)}.  }
\thanks{ J. Zhang, L. Zhang and B. Zou are with the Department of Information Engineering, Harbin Institute of Technology, Harbin, China. (e-mail: lmzhang@hit.edu.cn; 24b905016@stu.hit.edu.cn)}

}

\markboth{Journal of \LaTeX\ Class Files,~Vol.~14, No.~8, August~2021}%
{Shell \MakeLowercase{\textit{et al.}}: A Sample Article Using IEEEtran.cls for IEEE Journals}

\maketitle

\begin{abstract}
Synthetic Aperture Radar (SAR) image captioning enables scene-level semantic understanding and plays a crucial role in applications such as military intelligence and urban planning, but its development is limited by the scarcity of high-quality datasets. To address this, we present FSAR-Cap, a large-scale SAR captioning dataset with 14,480 images and 72,400 image–text pairs. FSAR-Cap is built on the FAIR-CSAR detection dataset and constructed through a two-stage annotation strategy that combines hierarchical template-based representation, manual verification and supplementation, prompt standardization. Compared with existing resources, FSAR-Cap provides richer fine-grained annotations, broader category coverage, and higher annotation quality. Benchmarking with multiple encoder–decoder architectures verifies its effectiveness, establishing a foundation for future research in SAR captioning and intelligent image interpretation. The dataset will be publicly available at: \url{https://github.com/hitjiao/FSAR-Cap}

\end{abstract}


\section{Introduction}
\IEEEPARstart{S}{ynthetic} Aperture Radar (SAR) is an active imaging system that provides all-weather, day-and-night observation capabilities, making it highly valuable for applications such as disaster monitoring, urban planning, and military intelligence \cite{dong1,my}. With the rapid development of deep learning, traditional image interpretation tasks—including classification, detection, and recognition—have achieved notable progress. Nevertheless, these methods primarily emphasize local features, such as individual targets or regions, and often fail to capture the comprehensive semantic relationships inherent in complex SAR scenes. Furthermore, they continue to rely heavily on expert knowledge and manual intervention, which hinders progress toward fully automated SAR image interpretation.

In contrast, image captioning has emerged as a novel paradigm for SAR image interpretation, effectively bridging the gap between visual data and natural language. This approach not only identifies objects within SAR imagery but also generates semantically coherent textual descriptions that reflect the overall scene. By shifting from the traditional focus on “points, lines, and surfaces” toward a “scene-level” understanding, image captioning facilitates higher-level semantic comprehension and significantly enhances the degree of automation. In practical scenarios such as emergency response, military reconnaissance, and fine-grained urban management, it holds great potential as a decision-support tool.

Nevertheless, the intrinsic characteristics of SAR imaging—such as blurred object contours, incomplete structural information, and low target-to-background contrast—make semantic extraction and representation highly challenging. Addressing this issue requires the construction of large-scale, high-quality SAR captioning datasets. Several datasets have been developed to date. SSICD \cite{ssicd} and HRSSRD-Captions \cite{ship} were among the first manually annotated SAR captioning datasets; however, they were confined to ship-related categories, limiting their general applicability. SARChat \cite{sarchat} represents a large-scale SAR–text dataset, but its rigid, template-based annotations constrain descriptive flexibility. SARLANG \cite{SARLANG} utilizes paired optical–SAR imagery for annotation, yet the resulting captions tend to be short and of relatively low quality. ATRNet-SARCap \cite{air} introduced an innovative semi-supervised hierarchical framework based on GPT-4V, though its coverage was limited to aircraft and related scenes. More recently, SAR-TEXT \cite{SAR-TEXT} has been proposed as a large-scale captioning dataset constructed from object detection, semantic segmentation, and optical–SAR paired data. However, template-based annotations remain overly rigid, while direct annotation with large vision–language models (e.g., GPT) often results in quality inconsistencies. Furthermore, most existing datasets cover only limited categories and lack detailed, fine-grained annotations, restricting both the expressiveness and the quality of captions.

To mitigate the scarcity of high-quality SAR captioning datasets and the lack of fine-grained annotations, we propose a two-stage annotation framework. First, object detection annotations are utilized to generate template-based descriptions, encompassing 10 scene types and 25 distinct templates to produce both image-level and object-level captions. Second, human annotators perform manual verification and supplementation to enrich semantic content. Subsequently, optimized prompts are employed to refine and standardize the final annotations. FAIR-CSAR \cite{FAIR-CSAR}, the most fine-grained SAR object detection dataset to date, serves as the foundation for this work, covering 22 object categories across airports, ports, and riverine regions in 32 global locations. Building upon this foundation, we construct FSAR-Cap, the first large-scale SAR captioning dataset with fine-grained annotations. FSAR-Cap comprises 14,480 images paired with 72,400 image–text annotations. Furthermore, we conduct systematic modeling and empirical evaluation using mainstream encoder–decoder architectures to demonstrate the effectiveness of FSAR-Cap in advancing SAR captioning performance.

The main contributions of this work are summarized as follows:
\begin{enumerate}
    \item We propose a two-stage annotation framework that integrates template-based representation, manual supplementation, prompt standardization, thereby significantly enhancing annotation quality.
    \item We construct FSAR-Cap, the first large-scale SAR captioning dataset with fine-grained annotations, comprising 72,400 image–text pairs.
    \item We conduct comprehensive evaluations of multiple encoder–decoder architectures and state-of-the-art image captioning models on FSAR-Cap, establishing a strong foundation for future research on SAR-specific captioning models.
\end{enumerate}

\begin{figure*}[t]
    \centering
    \includegraphics[width=1\linewidth]{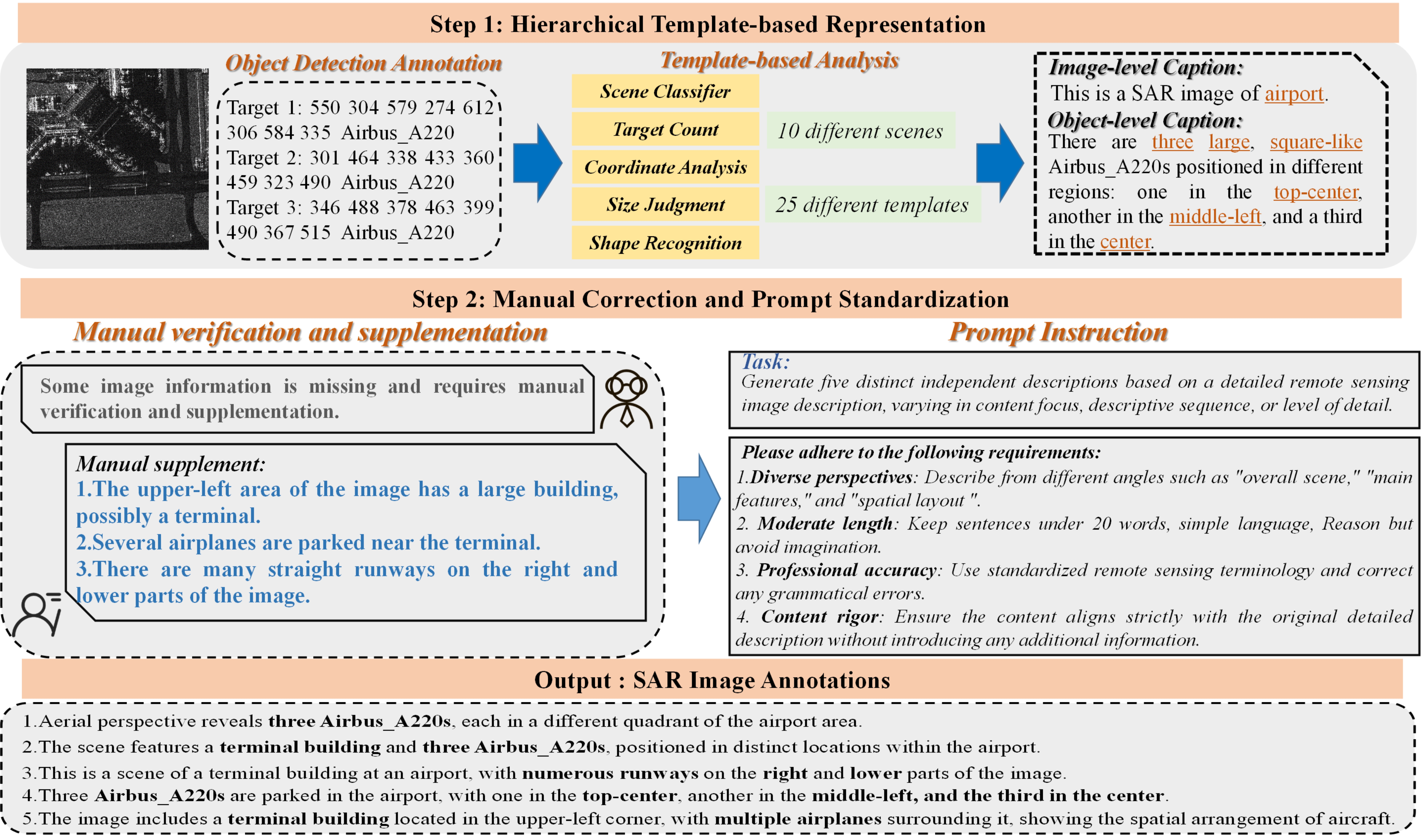}
    \caption{Two-stage annotation strategy: Stage 1 generates image-level and object-level captions via hierarchical template-based representation; Stage 2 refines the annotations through manual correction, supplementation, and prompt standardization.}
    \label{caption}
\end{figure*}

\section{Methodology}
\subsection{Overall Framework}
The proposed two-stage annotation framework is illustrated in Fig. \ref{caption}. In the first stage, based on the FAIR-CSAR object detection dataset, a template-based representation is employed for annotation generation. Specifically, a classifier covering 10 scene categories is trained to provide image-level captions. To avoid excessive rigidity in the generated templates, 25 distinct templates are designed to perform target counting, coordinate analysis, size estimation, and shape recognition, thereby generating object-level captions. In the second stage, manual verification and supplementation are conducted to enrich the descriptive content. Finally, by applying carefully designed prompts, information from all components is integrated to produce five distinct captions for each image.

\subsection{Hierarchical Template-based Representation}
To address the aforementioned challenges, we propose a hierarchical template-based representation. To maximize the use of annotation information and enhance template diversity, we design 25 distinct templates that perform target counting, coordinate analysis, size estimation, and shape recognition, thereby generating object-level semantic annotations.

Target counting is implemented by statistically analyzing the occurrence frequency of object categories within the detection dataset. For categories containing fewer than ten instances, specific numerical expressions are used, whereas categories with more than ten instances are described using quantifiers such as “many” or “a lot of.” Coordinate analysis is uniformly encoded using the bounding box format ${<x_1, y_1><x_2, y_2><x_3, y_3><x_4, y_4>}$, from which the center coordinates are computed to determine the spatial position of each object within the image. The spatial relationships among objects are organized according to a standard 3×3 grid system, comprising nine regions: top-left, top-center, top-right, middle-left, center, middle-right, bottom-left, bottom-center, and bottom-right. Furthermore, the bounding box coordinates allow computation of object width and height, enabling both size estimation and shape recognition based on these parameters. The corresponding computational formulas are as follows:
\begin{equation}
\begin{aligned}
 SJ = \frac{w_{\text{box}} \times h_{\text{box}}}{W_{\text{img}} H_{\text{img}}} \times 100\% \quad\quad\quad  SR= \frac{h_{\text{box}}}{w_{\text{box}}}  
\end{aligned}
\end{equation}

Where $w_{box}$ and $h_{box}$ denote the width and height of the target bounding box, respectively, and $W_{img}$ and $H_{img}$ represent the width and height of the image. SJ (Size Judgment) represents the relative size of the object within the image: \textit{small} if $SJ < 0.5\%$, \textit{large} if $0.5\% < SJ < 15\%$, and \textit{huge} if $SJ > 15\%$. SR (Shape Recognition) represents the object's shape: \textit{square-like} if $0.7 < SR < 1.5$, and \textit{elongated} if $SR < 0.7$ or $SR > 1.5$.

To preserve global contextual information, a scene-level classifier was trained to recognize 10 scene categories (e.g., industrial area, river, airport, and port). The following template was then applied to generate image-level annotations: \textit{``This is a SAR image of \textless scene \textgreater.''} Thus, the generated annotations also include image-level semantic information.
\begin{figure*}[t]
    \centering
    \includegraphics[width=1\linewidth]{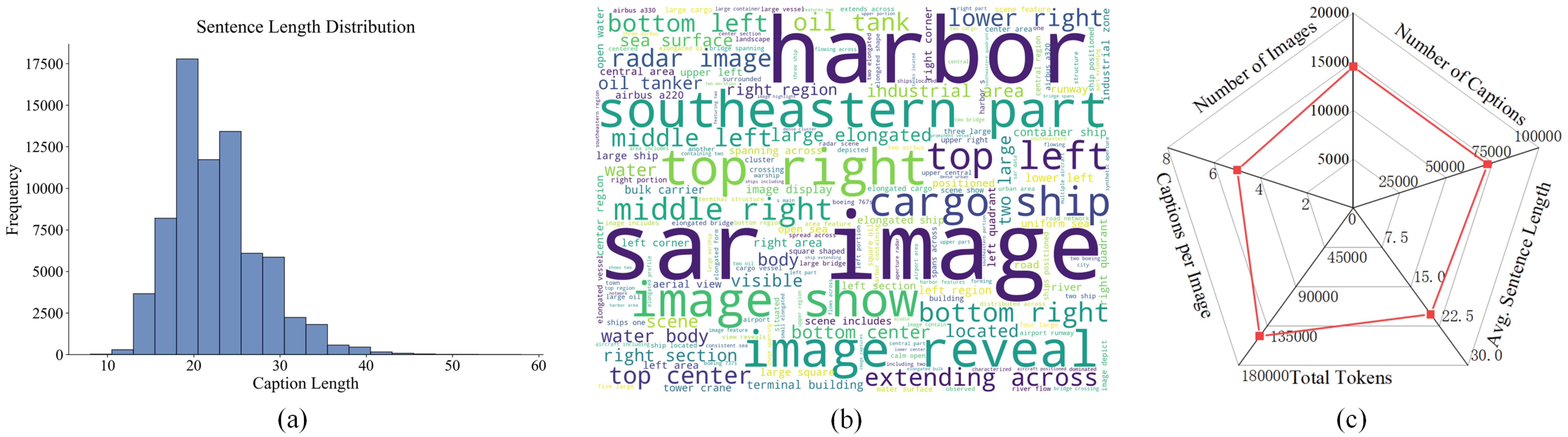}
    \caption{Statistics of the FSAR-Cap dataset. (a) Frequency histogram of the captions length per image. (b) Word cloud image of the primary annotation words in the dataset. (c) Radar plot of the main data in the FSAR-Cap dataset.}
    \label{mapgf}
\end{figure*}

\begin{table*}[!t]
    \centering
    \caption{Comparison of SAR Image Captioning Datasets}
    \label{tab:sar_caption_datasets}
    \resizebox{0.98\textwidth}{!}{
    \begin{tabular}{lccccp{3.2cm}p{4.5cm}}
    \toprule
    \textbf{Dataset} & \textbf{Images} & \textbf{Captions} & \textbf{Fine-grained} & \textbf{Annotation method} & \textbf{Target categories} \\
    \midrule
    SSIDC\cite{ssicd} & 1,500 & 7,500 & × & Manual annotation & Ship \\
    HRSSRD-Cap\cite{ship} & 1,000 & 5,000 & × & Manual annotation & Ship \\
    SARChat-Cap\cite{sarchat} & -- & 52,173 & × & Template & 5 categories \\
    SARLANG-Cap\cite{SARLANG} & 13,346 & 45,650 & × & Optical-SAR paired images & -- \\
    SAR-TEXT\cite{SAR-TEXT} & -- & 136,584 & × & Template + RS-Captioner & -- \\
    ATRNet-SARCap\cite{air} & 5,251 & 47,259 & × & GPT-4V & Aircraft \\
    FSAR-Cap & 14,480 & 72,400 & \checkmark & Template + Manual + LLM  & 22 subcategories \\
    \bottomrule
    \end{tabular}
    }
\end{table*}

\subsection{Manual Correction and Prompt Standardization}
In Step 1, annotations were performed from both image-level and object-level perspectives. However, further analysis revealed that the annotated information was still incomplete and failed to comprehensively capture the overall semantics of SAR images. As illustrated in Fig. \ref{caption}, although the model correctly identified the scene as an airport, it did not recognize critical features such as the terminal building in the upper-left corner or the aircraft runways on the right and lower sides. Such fine-grained details are often challenging for existing models to detect automatically. Furthermore, when using current vision–language models (VLMs) for annotation, the lack of dedicated training on SAR data can easily result in inaccurate or misleading descriptions.

To address these limitations, a manual verification and supplementation process was introduced in the second stage to enhance the semantic integrity and accuracy of the generated annotations. During this stage, human annotators meticulously reviewed the automatically generated outputs to ensure semantic correctness, linguistic coherence, and descriptive diversity. Missing or ambiguous details—such as object attributes, spatial relationships, and scene-level semantics—were manually refined and supplemented, thereby improving the completeness and reliability of the final annotations.

Following verification, all information components were integrated using carefully designed prompt templates, which guided the model to organize and articulate textual content in a natural, coherent, and semantically consistent manner. Ultimately, five semantically distinct captions were generated for each image, each highlighting different aspects of the scene, objects, and spatial configurations, thus producing a comprehensive, diverse, and semantically rich caption set.

\subsection{FSAR-Cap Dataset Analysis}
The proposed dataset, FSAR-Cap, is the first large-scale SAR captioning dataset with fine-grained textual annotations. It comprises 14,480 images and 72,400 image–text pairs. To ensure annotation quality, a comprehensive dataset analysis was conducted, as illustrated in Fig. \ref{mapgf}. Fig. \ref{mapgf}(a) presents the frequency histogram of caption lengths per image, indicating that most captions contain between 20 and 25 words, which reflects the richness of the annotations. Fig. \ref{mapgf}(b) shows the dataset’s word cloud, where fine-grained category terms such as “cargo ship” and “Boeing 747”, as well as spatial expressions like “top left” and “bottom”, appear frequently. Fig. \ref{mapgf}(c) provides a radar chart offering an overview of annotation quality. The average caption length is 20.8 words, with a total of approximately 1.5 million tokens across the dataset.

A comparison of existing SAR captioning datasets is summarized in Table \ref{tab:sar_caption_datasets}. FSAR-Cap surpasses previous datasets in terms of scale, category diversity, and sentence length. Notably, it is the first SAR captioning dataset to provide fine-grained, semantically rich annotations.

\section{Experiment}
\subsection{Experimental Setting}
\begin{figure}[h]
    \centering
    \includegraphics[width=1\linewidth]{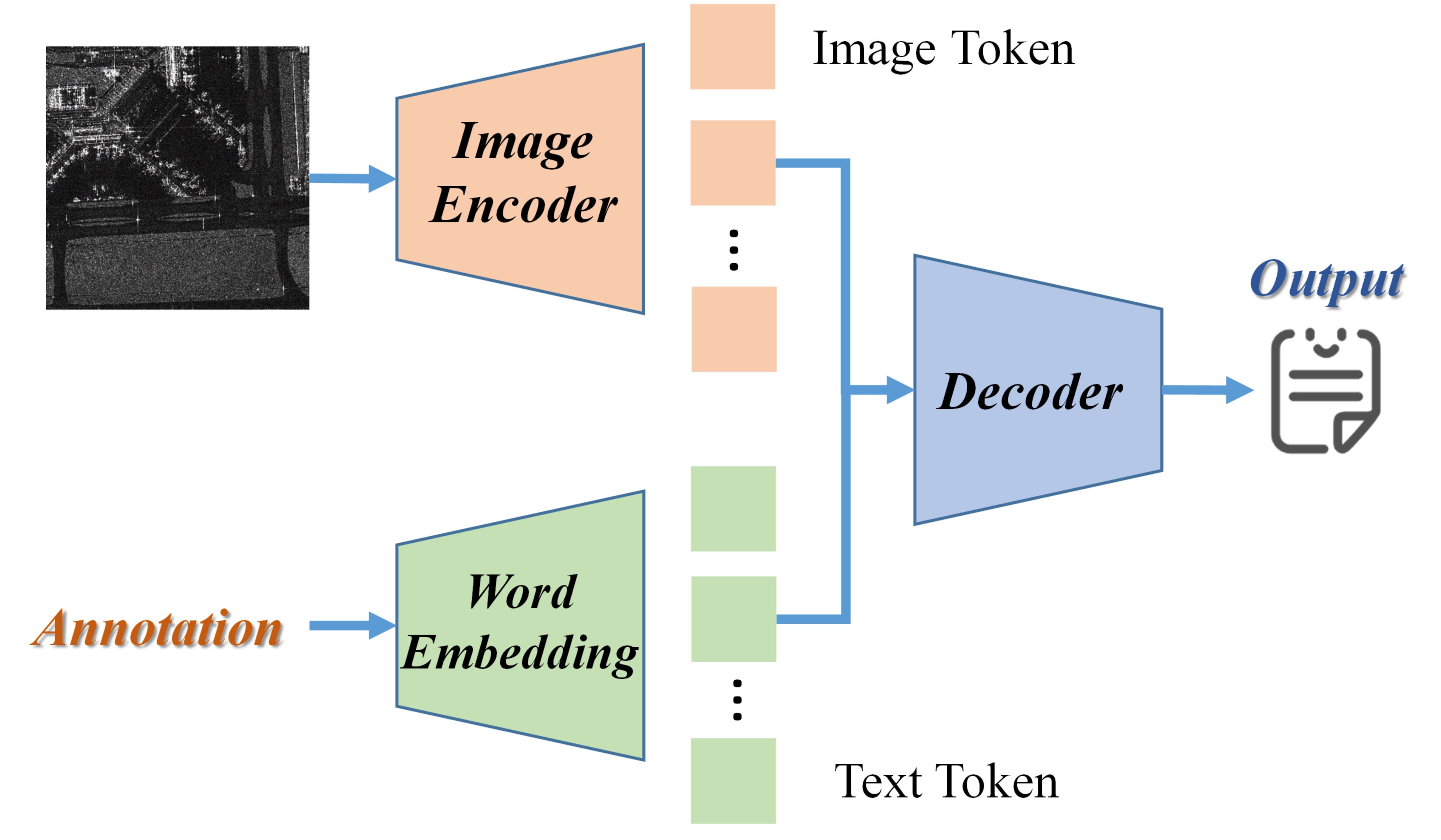}
    \caption{Overview of the encoder–decoder architecture employed for SAR image captioning.}
    \label{encoder}
 \end{figure}

To assess the effectiveness of the proposed FSAR-Cap dataset for image–language modeling, experiments were conducted on the SAR image captioning task using a mainstream encoder–decoder framework, as illustrated in Fig. \ref{encoder}. The framework consists of two components: feature encoding and caption decoding. The encoder extracts high-level visual features from SAR images to generate semantic representations, using several popular backbones, including VGG16, VGG19, ResNet50, ResNet101, ViT, and ConvNeXt. The decoder integrates visual and textual tokens through word embedding to produce semantically aligned and fluent captions. Two decoder architectures, LSTM and Transformer, were evaluated. During training, the model minimizes cross-entropy loss to reduce discrepancies between generated and reference captions, ensuring accuracy and fluency. Moreover, we compared the performance of representative image captioning models, including Soft-Attention \cite{attention}, Hard-Attention \cite{attention}, FC-Att \cite{Att}, SM-Att \cite{Att}, MLCA \cite{MLCA}, MLAT \cite{MLAT}, HCNet \cite{HCNet}, and PureT \cite{PureT}.

For the experimental setup, the dataset was partitioned into 10,000 images for training, 2,240 for validation, and 2,240 for testing. All encoder networks were initialized with weights pre-trained on ImageNet, and input images were resized to 224×224. The decoder embedding dimension was set to 512 with three layers. Models were trained for 100 epochs with a batch size of 64, and early stopping was applied if no improvement was observed for five consecutive epochs. The Adam optimizer was employed with a learning rate of 1×10$^{-4}$. During inference, beam search with a beam size of 5 was used. All experiments were conducted on a workstation equipped with an NVIDIA RTX A6000 GPU.

Model performance is evaluated based on the similarity between generated and reference captions. Common automatic evaluation metrics include \textit{BLEU}, \textit{METEOR}, \textit{ROUGE-L}, \textit{CIDEr} and \textit{$S$}, each assessing sentence similarity from a distinct perspective. Higher scores on these metrics indicate greater consistency and accuracy in the generated captions.

\subsection{Performance Comparisons}
To evaluate the effectiveness of the FSAR-Cap dataset constructed through the proposed two-stage annotation strategy, we conducted a series of comparative experiments using various encoder–decoder architectures and several state-of-the-art image captioning models. The corresponding experimental results are presented in Tables \ref{tab:encoder_decoder_comparison} and \ref{tab:comparison}.

\begin{table}[h!]
\centering
\caption{Performance of different encoders and decoders on the FSAR-Cap dataset. BLEU-1, BLEU-4, METEOR, ROUGE-L, CIDEr, and S are denoted as B1, B4, M, R, C, and S, respectively. The best-performing results are highlighted in bold.}
\setlength{\tabcolsep}{3pt} 
\renewcommand{\arraystretch}{1} 
\begin{adjustbox}{max width=\textwidth}
\begin{tabular}{c|c |c c c c c c}
\toprule
\textbf{Encoder} & \textbf{Decoder} & \textbf{B1} & \textbf{B4} & \textbf{M} & \textbf{R} & \textbf{C} & \textbf{S} \\
\midrule
\textit{VGG16} & \multirow{6}{*}{\textit{LSTM}} & 55.39 & 23.65 & 22.95 & 44.02 & 40.10 & 32.68 \\
\textit{VGG19} &  & 57.49 & 25.16 & 24.98 & 46.27 & 46.43 & 35.71 \\
\textit{ResNet50} &  & 65.44 & 30.58 & 25.41 & 47.63 & 64.88 & 42.12 \\
\textit{ResNet101} &  & 63.94 & 29.48 & 24.91 & 47.16 & 61.97 & 40.88 \\
\textit{VIT} &  & 60.28 & 27.88 & 23.67 & 46.48 & 53.98 & 38.00 \\
\textit{ConvNext} &  & 59.97 & 27.60 & 23.43 & 46.03 & 53.80 & 37.71 \\
\midrule
\textit{VGG16} & \multirow{6}{*}{\textit{Transformer}} & 68.19 & 34.85 & 26.36 & 50.33 & 82.61 & 48.54 \\
\textit{VGG19} &  & 68.45 & 35.53 & 26.84 & 50.69 & 83.06 & 49.03 \\
\textit{ResNet50} &  & \textbf{70.50} & 36.82 & 27.98 & 51.83 & 88.46 & 51.27 \\
\textit{ResNet101} &  & 69.87 & \textbf{37.92} & \textbf{28.25} & \textbf{52.78} & \textbf{90.22} & \textbf{52.30} \\
\textit{VIT} &  & 69.34 & 36.47 & 27.74 & 51.83 & 83.97 & 50.00 \\
\textit{ConvNext} &  & 68.14 & 34.62 & 26.85 & 50.50 & 82.89 & 48.72 \\
\bottomrule
\end{tabular}
\end{adjustbox}
\label{tab:encoder_decoder_comparison}
\end{table}

As shown in Table \ref{tab:encoder_decoder_comparison}, the results obtained from different encoder–decoder configurations on FSAR-Cap reveal that each encoder architecture exhibits distinct advantages in feature extraction and semantic representation. At the same time, the choice of decoder significantly affects the fluency and overall quality of the generated captions. Notably, the Transformer-based decoder consistently outperforms its LSTM-based counterpart. When ResNet101 is used as the encoder in combination with a Transformer decoder, the model achieves superior performance across all evaluation metrics, including BLEU-4, METEOR, ROUGE-L, CIDEr, and SPICE.

\begin{table}[h!]
\centering
\caption{Performance of different image captioning models on the FSAR-Cap dataset.}
\setlength{\tabcolsep}{5pt}
\renewcommand{\arraystretch}{1}
\begin{adjustbox}{max width=\textwidth}
\begin{tabular}{c|cccccc}
\toprule
\textbf{Methods} & \textbf{B1} & \textbf{B4} & \textbf{M} & \textbf{R} & \textbf{C} & \textbf{S} \\
\midrule
\textit{Soft-attention}\cite{attention} & 68.75 & 36.21 & 27.26 & \textbf{51.86} & 78.88 & 48.55 \\
\textit{Hard-attention}\cite{attention} & 63.61 & 31.97 & 25.10 & 48.46 & 62.57 & 42.03 \\
\textit{FC-Att}\cite{Att} & 66.55 & 32.94 & 26.39 & 49.44 & 75.82 & 46.15 \\
\textit{SM-Att}\cite{Att} & 67.21 & 32.38 & 26.35 & 49.18 & 71.33 & 44.81 \\
\textit{MLCA}\cite{MLCA} & 67.64 & 32.66 & 26.63 & 49.27 & 72.35 & 45.23 \\
\textit{MLAT}\cite{MLAT} & 68.84 & \textbf{36.68} & 27.80 & 51.79 & \textbf{84.54} & \textbf{50.20} \\
\textit{HCNet}\cite{HCNet} & \textbf{70.95} & 35.89 & 28.08 & 51.59 & 79.90 & 48.87 \\
\textit{PureT}\cite{PureT} & 65.70 & 33.05 & \textbf{28.86} & 49.53 & 71.04 & 45.62 \\
\bottomrule
\end{tabular}
\end{adjustbox}
\label{tab:comparison}
\end{table}

The results of various mainstream image captioning models on FSAR-Cap are summarized in Table \ref{tab:comparison}. Models such as Soft-Attention, Hard-Attention, FC-Att, and SM-Att, which integrate attention mechanisms within the LSTM framework, yield substantial improvements over the LSTM baseline in Table \ref{tab:encoder_decoder_comparison}, confirming that attention mechanisms enhance a model’s ability to focus on salient targets within the scene. Furthermore, models such as MLAT and HCNet, originally developed for optical remote sensing image captioning, also demonstrate competitive performance, suggesting that methods designed for optical imagery retain partial applicability to SAR image captioning. Nevertheless, no existing captioning model is specifically tailored for SAR imagery, which fails to address unique SAR characteristics such as speckle noise, scattering properties, and imaging mechanisms. Accounting for these characteristics is essential for further advancing SAR image captioning performance.

\subsection{Case Study}
\begin{figure}[h]
    \centering
    \includegraphics[width=1\linewidth]{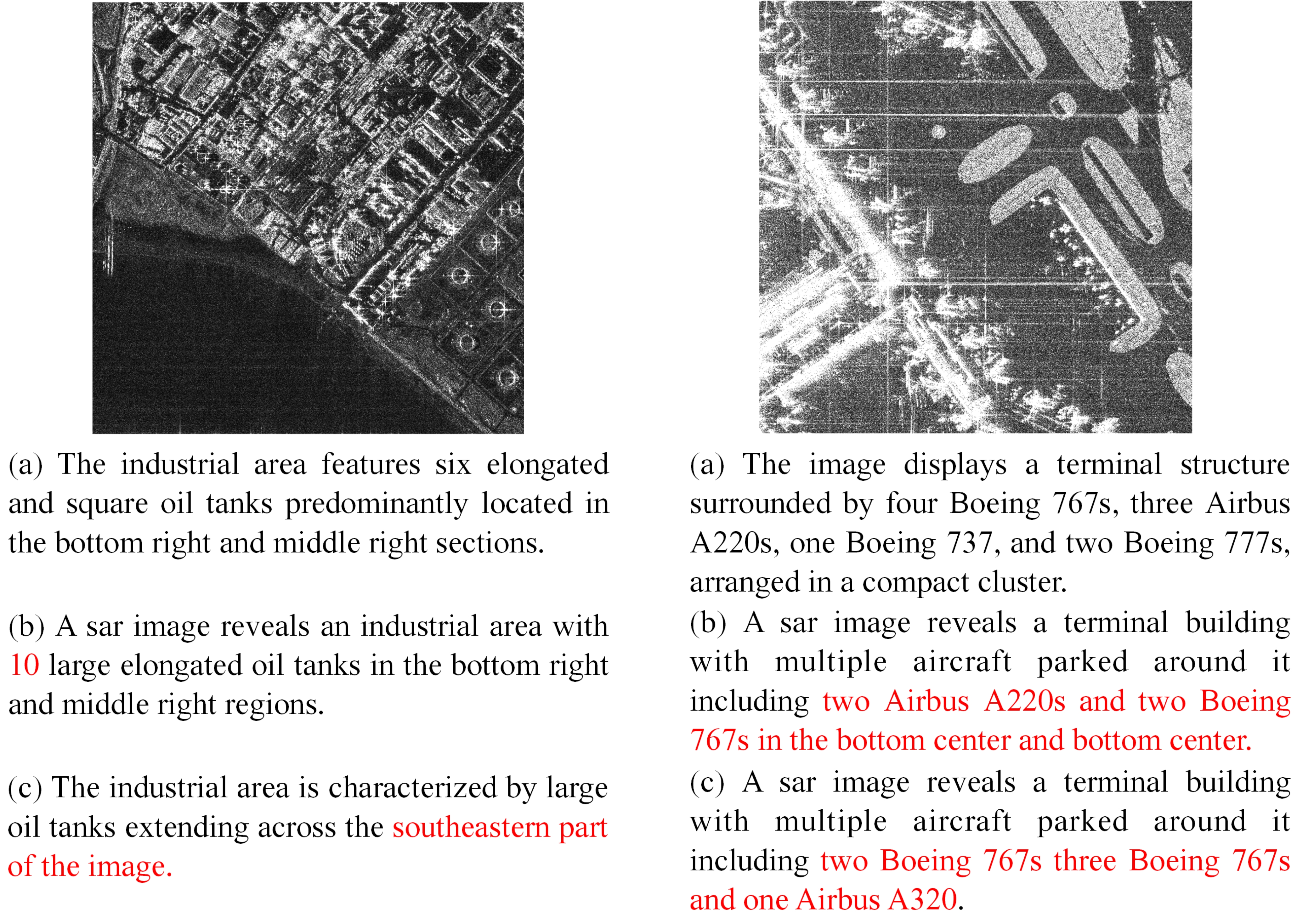}
    \caption{Image captioning case study. (a) One of the five ground-truth captions. (b) Generated by the ResNet101-Transformer. (c) Generated by the MLAT. Red words mark parts that don not match the images.}
    \label{case}
 \end{figure}

To illustrate the captioning performance of the models on the FSAR-Cap dataset, we present the results of ResNet101-Transformer and MLAT on two sample images, as shown in Fig. \ref{case}. In the first image, both models correctly identify the region as an industrial area, but neither can accurately count the number of oil tanks, and MLAT makes an error in spatial localization. In the second image, both models successfully describe the airport and terminal and detect multiple aircraft, although they misestimate the number of different aircraft types.
 
Overall, the models perform well in scene recognition and spatial localization but show limitations in object counting and fine-grained category recognition. Nonetheless, the generated captions effectively capture most of the key semantic content in the images, demonstrating the effectiveness of the proposed annotation method.

\section{conclusion}
In summary, we propose FSAR-Cap, the first large-scale SAR captioning dataset featuring fine-grained textual annotations. Experimental results across multiple encoder–decoder architectures and mainstream image captioning methods demonstrate its effectiveness and potential to advance automatic SAR interpretation. In future work, FSAR-Cap can serve as a valuable benchmark for the community. We anticipate the emergence of captioning models specifically tailored for SAR imagery, and the dataset may further facilitate the development of more robust and generalizable VLMs in the SAR domain.

\bibliographystyle{IEEEtran}
\bibliography{manuscript.bib}

\end{document}